# Field emission: Why converting LAFE voltages to macroscopic fields before making a Fowler-Nordheim plot has often led to spurious characterization results

Richard G. Forbes[a]

*Advanced Technology Institute & Dept. of Electrical and Electronic Engineering, University of Surrey, Guildford, Surrey GU2 7XH, United Kingdom*

An important parameter used to characterize large-area field electron emitters (LAFEs) is the characteristic apex field enhancement factor $\gamma_C$. This parameter is normally extracted from the slope of a Fowler-Nordheim (FN) plot. Several years ago, the development of an "orthodoxy test" allowed a sample of 19 published FN plots relating to LAFEs to be tested, and it was found that about 40% of the related papers were reporting spuriously high values for $\gamma_C$. In technological papers relating to LAFE characterization, common practice is to pre-convert the measured voltage into an (apparent) value of macroscopic field before making and analyzing a FN plot. This paper suggests that the cause of the "spurious-FEF-value" problem is the widespread use of a pre-conversion equation that is *defective* (for example, not compatible with ordinary electrical circuit theory) when it is applied to so-called "non-ideal" field emission devices/systems. Many real devices/systems are non-ideal. The author argues that FN plots should be made using raw experimental current-voltage data, and that an orthodoxy test should be applied to the resulting FN plot before any more-detailed analysis, and that (in view of growing concerns over the reliability of published "scientific" results) reviewers should scrutinize field emission materials-characterization papers with enhanced care.

[a]Electronic mail: r.forbes@trinity.cantab.net

## I. INTRODUCTION

### A. Background

Large-area field electron emitters (LAFE's) typically have a "footprint area" of between 1 mm$^2$ and 1 cm$^2$ and comprise very many individual emitters. LAFEs have various potential technological applications as large-area electron sources[1-3], and there has been much interest in exploring the effectiveness of different fabrication materials and procedures.

A common characterization procedure is to place the LAFE on one of a pair of well-separated planar parallel plates and make current-voltage $I_m(V_m)$ measurements. There has been much interest in characteristic values of *apex field enhancement factors* (apex FEFs). In field electron emission (FE) literature, FEFs are normally denoted by the symbol $\beta$, but to avoid ambiguities with other uses of the symbol $\beta$ in FE literature, FEFs are denoted here by $\gamma$.

For a given emitter, in the planar-parallel-plate (PPP) geometry, its true apex FEF $\gamma_{Pa}^{true}$ is defined by

$$\gamma_{Pa}^{true} = F_a/F_P^{true}, \tag{1}$$

where $F_a$ is the absolute magnitude of the local electrostatic field at the emitter apex, and the "true plate field" $F_P^{true}$ is the true value of the absolute magnitude of the mean electrostatic field between the plates. $F_P^{true}$ is related adequately to the (positive) voltage $V_P$ applied to the anode plate, relative to the cathode plate, by

$$F_P^{true} \approx V_P/d_P , \tag{2}$$

where $d_P$ is the separation of (the inwards facing surfaces of) the plates. This formula is strictly correct only if the plates are smooth and flat, and have equal uniform work function, but it is a good enough approximation for virtually all working purposes.

For a LAFE, there is particular interest in the apex FEFs for the most strongly emitting individual sites. In important respects, these characterize the LAFE. The corresponding apex field is denoted here by $F_C$ and the corresponding true apex FEF by $\gamma_{PC}^{true}$.

A matter of interest is the relationship between $F_C$ and the measured voltage $V_m$ that is applied to an FE device/system by a high-voltage generator (HVG), or equivalent. This relationship can be written in the form

$$F_C = V_m/\zeta_C , \tag{3}$$

where $\zeta_C$ is called here the *characteristic voltage conversion length (VCL)*. Since both $F_C$ and $V_m$ are in principle well defined, it follows that $\zeta_C$ is also a well-defined parameter. $\zeta_C$ is not a physical length, except in special geometrical circumstances. Rather, it is a system characterization parameter: when $\zeta_C$ is relatively small, the emitter turns on at a relatively low voltage.

A FE device/system is described as *ideal* if there are no "complications" as described below and its measured emission characteristics are determined only by the geometry of the system and the physics of emission from a surface that has fixed unchanging shape and a work function that does not vary significantly with local surface field or with emission current density. Data and other properties that relate to ideal devices/systems are also described here as "ideal". For an ideal device/system the VCL $\zeta_C$ is *constant*, independent of the measured current and/or voltage.

However, often real systems exhibit "complications", such as (amongst others) leakage currents, series resistance in the current path to the HVG, current dependence in field enhancement factors, field-dependent changes in emitter geometry, space-charge effects, field-penetration effects (non-metals only), and reversible gas-adsorption effects. "Complications" can cause "non-ideality" by which the VCL $\zeta_C$ becomes a function of measured current and/or voltage, and Fowler-Nordheim (FN) plots made by assuming that the device/system is ideal are distorted and defective.

## B. The traditional current/voltage FN plot

The traditional method[4] of analyzing FE $I_m(V_m)$ data has been to make the corresponding $I_m(V_m)$-type FN plot, that is, a plot of the form $\ln\{I_m/V_m^2\}$ vs $1/V_m$ (or equivalent using logarithms to base 10, but we consider natural logarithms here). Murphy-Good (1956) FE theory[5] (which corrected physical mistakes made in the derivation[6] of the original 1928 Fowler-Nordheim equation) predicts that theoretical FN plots should be *approximately* straight; thus, it is predicted that it will be possible to fit to experimental data points in a FN plot a straight line that has slope $S_V^{fit}$ and intercept $\ln\{R_V^{fit}\}$. For ideal data, an extracted VCL value $\zeta_C^{extr}$ can be obtained from the formula (see Appendix)

$$\zeta_C^{extr} = - S_V^{fit} / s_t b \phi^{3/2} , \qquad (4)$$

where $b$ is the second FN constant[7], $\phi$ is the local work function, and $s_t$ is a slope correction factor that can usually be adequately approximated as 0.95.

From eqns (1) and (2), it can be shown that

$$\gamma_{PC}^{true} = (d_P/\zeta_C)(V_m/V_P) . \qquad (5)$$

An ideal PPP device/system has $V_P = V_m$, so (with ideal data) an extracted FEF-value can be obtained from the formula

$$\{\gamma_{PC}\}^{extr} = d_P/\zeta_C^{extr} = -(d_P s_t b \phi^{3/2})/S_V^{fit}. \tag{6}$$

A so-called "orthodoxy test" has been devised[8], in order to test whether a particular FN data plot is ideal. The basic idea is as follows. A FN plot can be used to deduce the range of local field ($F_C$) and scaled-field ($f$) [see eq. (A3)] values *apparently* involved in the measurements under test. If the deduced field values (or some of them) are greater than the known local fields at which field emitters are known to melt, explode or otherwise self-destruct, then it may reasonably be concluded that the FN plot is defective/distorted and that any characterization-parameter values extracted from it are likely to be spurious. (The actual test, which is applied using a spreadsheet, is somewhat more sophisticated than this, is a form of "engineering triage test", and is robust.)

It needs to be made clear that there is nothing intrinsically unsatisfactory about non-ideal FE devices/systems: some such devices/systems are technologically useful when it is required to "ballast" an emitter in order to inhibit high-current run-away (e.g., ref. 9). The objection is to *mis-characterization* of non-ideal FE devices/systems.

## C. The modern current-density/macroscopic-field FN plot

Notwithstanding the traditional approach just described, most modern technological literature does the analysis of experimental FE $I_m(V_m)$ data in a slightly different fashion that involves *pre-conversion* of the measured $I_m(V_m)$ data into a different format, before making a FN plot. For PPP-geometry devices/systems, this is usually done in the following way. Measured current values are pre-converted to values of the macroscopic ("LAFE-average") emission current density $J_{M,m}$ by using the pre-conversion equation:

$$J_{M,m} = I_m / A_M, \tag{7}$$

where $A_M$ is the LAFE macroscopic area (or "footprint"). $A_M$ is a well-defined measurable parameter, and there is no problem with this conversion, provided the value of $A_M$ is published. Measured voltage values are pre-converted to apparent values $F_P^{app}$ of the plate field by using the pre-conversion equation:

$$F_P^{app} = V_m/d_P. \tag{8}$$

A FN plot of form $\ln\{J_{M,m}/(F_P^{app})^2\}$ vs $1/F_P^{app}$ is then constructed, and its slope $S_{FP}^{fit}$ determined.

An ideal plot of this form would have a predicted slope $S_{FP}^{theor}$ given by

$$S_{FP}^{theor} = -s_t b \phi^{3/2} / \gamma_{PC}^{true}. \qquad (9)$$

Hence, when $S_{FP}^{fit}$ is identified with $S_{FP}^{theor}$, an extracted "apparent" value $\gamma_{PC}^{app}$ of apex FEF can be derived from the equation

$$\gamma_{PC}^{app} = \gamma_{PC}^{extr} = -s_t b \phi^{3/2} / S_{FP}^{fit}. \qquad (10)$$

The orthodoxy test noted above can be applied to any form of FN plot. In practice, although only a relatively small sample (19) of papers was tested in ref. 8, it was found that around 40% failed the test, and were thus considered to be reporting spuriously high values for characteristic apex-FEF values.

This high incidence of spurious FEF values has not just been a matter of academic interest. Some years ago, the European Space Agency (ESA) put out a tender request, which would have had the effect of requiring the contractor to search the research literature for high apex-FEF values, in order that materials could be selected for practical research into the possibility of charge-neutralizing satellites in Earth-orbit, via field electron emission. The presumed high incidence of spuriously high apex-FEF values in the literature meant that probably the materials with the most spuriously high FEF-values would have been identified for further investigation by ESA. The orthodoxy test (and, later, the procedure[10] of "phenomenological adjustment") were originally developed as engineering responses to this unacceptable situation.

The purpose of the present note is point out the probable physical reason for the deduction and publication of spuriously high apex-FEF values. This is that the pre-conversion equation (8) is *defective* when the device/system is non-ideal. The related theory depends on the precise origin of the non-ideality. To illustrate the general nature of the problem, a specific potential cause of non-ideality will be considered, namely series resistance in the current path to the high-voltage generator. The author does not consider that this is usually the cause of observed non-ideality (though it appears to be in some cases—see ref. 11 and also doi:10.13140/RG.2.2.14817.35683), but the related theory is especially clear and simple.

In what follows, we first indicate the nature of the mistake that has often been made in modern FN-plot analysis, and then discuss some of the implications.

## II. AN APPARENTLY COMMON MISTAKE IN MODERN FN-PLOT ANALYSIS

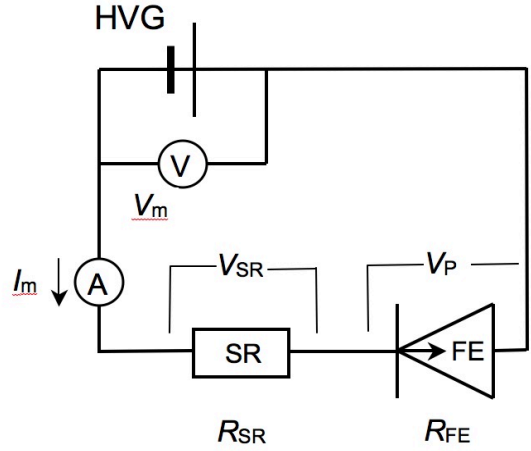

FIG. 1. Equivalent circuit for FE system comprising a resistance "SR" in series with a parallel-planar-plate (PPP) geometry field emitter "FE", with current $I_m$ driven by a high-voltage generator "HVG". The direction of electron flow, and all voltages $V$, are taken as positive. The resistances $R_{SR}$ and $R_{FE}$ are defined by $V_{SR}/I_m$ and $V_P/I_m$, respectively. The capacitance associated with the PPP geometry field emitter does not influence the steady-current values of interest here, and is not shown.

Consider an electrical-engineering-type equivalent circuit for the FE device/system as shown in Fig. 1. The measured voltage $V_m$ is the sum of the voltage $V_{SR}$ across the series resistance and the voltage $V_P$ between the parallel plates, thus:

$$V_m = V_P + V_{SR}. \qquad (11)$$

On dividing all terms in this equation by $d_P$, making use of eq. (2), and re-arranging, we get

$$F_P^{true} = (V_m/d_P) - (V_{SR}/d_P). \qquad (12)$$

What one needs, in order to construct a reliable FN plot involving plate fields, is the *true* plate-field. Thus, eq. (12), rather than eq. (8) above, would be the correct equation for pre-converting measured-voltage values, when non-ideality is caused by series resistance. If eq. (12) is written in the equivalent form

$$F_P^{true} = F_P^{app} - (V_{SR}/d_P), \qquad (13)$$

then it is obvious that the usual pre-conversion equation, namely eq. (8), has neglected the term

($V_{SR}/d_P$).

One can assume that most graduate students are totally familiar with rule (11) for addition of voltages in electrical circuit theory. It would seem that converting this rule into one about field values, by dividing by $d_{sep}$, has made the outcome so unfamiliar that many users can no longer get the underlying circuit theory correct. This perception has helped form the author's strong view that this aspect of FN-plot analysis is best carried out using voltages, since the likelihood of conceptual error is expected to be much less.

## III. DISCUSSION

### A. The immediate problem

For further discussion, it is helpful to put result (13) into a slightly different form. By trivial algebra, we have

$$F_P^{true} = \frac{V_m}{d_P} - \left(\frac{V_m}{d_P}\right)\left(\frac{d_P}{V_m}\frac{V_{SR}}{d_P}\right) = \frac{V_m}{d_P} - \left(\frac{V_m}{d_P}\right)\left(\frac{V_{SR}}{V_m}\right) = \frac{V_m}{d_P}\left(1 - \frac{V_{SR}}{V_m}\right). \quad (14)$$

And, on defining a "correction factor for series resistance" $\tau_{SR}$ by

$$\tau_{SR} = \left(1 - \frac{V_{SR}}{V_m}\right), \quad (15)$$

we can write

$$F_P^{true} = \tau_{SR}(V_m/d_P). \quad (16)$$

There are at least four possible causes of non-ideality that affect the relationships between fields and voltages. If these were all independent and all operating, then we would have

$$F_P^{true} = \tau_{SR}\tau_2\tau_3\tau_4 (V_m/d_P). \quad (17)$$

In reality, the theories of possible sources of non-ideality are not all well established, and the causes might interact. Thus, in practice, at present, it is *impossible in general* for an experimentalist to know in advance whether they can extract a true value of plate field (or of any other kind of macroscopic field) by pre-converting their measured voltages. [There are exceptions, such as devices/systems where the whole conduction path to the high-voltage generator is known to have very low resistance, and it is known that there are no effects operating that are equivalent to "field-dependent emitter geometry.]

At this point, there are three options open to experimentalists. First, disregard considerations of the type discussed here, and continue with well-established community practice. The disadvantage of this is the need to avoid the following situation, which limits the usefulness of the paper:

(a) a published paper does not contain the original (raw) experimental results; and
(b) the method by which the original data have been pre-converted is not described, and/or the value of a parameter used in pre-conversion (equivalent to $d_P$ here) is not stated; and
(c) application of the orthodoxy test to a published FN plot shows that the plot is defective and that any characterization parameter derived from it is highly likely to be spurious.

In the context of determining apex-FEF values, published results of this kind have been described[12] as "waste-bin" results, as far as the reader is concerned.

A second option is to present the pre-converted data, but also apply the orthodoxy test, and report the result. The disadvantage of this is that, if the FN plot fails the orthodoxy test, then the paper is acknowledging that the plotted data presented are defective/distorted.

A third option, strongly preferred by the author, is to make and present an $I_m(V_m)$-type FN plot (or alternatively, a $J_{M,m}(V_m)$-type FN plot—provided the value of $A_M$ is stated, or alternatively a Murphy-Good (MG) plot[13] of either of these types). An appropriate orthodoxy test should then be applied to this plot, and the result reported. If the test is passed, than a characteristic apex-FEF value can be derived from the extracted VCL value, by using formula (6) (or the equivalent formula for an MG plot). If the test is failed, then at least the paper contains an accurate report of the raw experimental data. Additionally, it may be possible to apply the technique[10] of "phenomenological adjustment" to extract some characterization information.

A further good reason for always using the measured voltage $V_m$ when making FN or MG plots is that consistent practice of this kind would reduce the number of slightly different forms of plot-interpretation theory that need to be set out in the literature, and in the longer term would reduce the overall complexity of the subject.

For definiteness and simplicity, the discussion here has assumed PPP geometry. However, other system geometries are also used in FE experiments, in particular systems where a needle-like "anode probe" is brought up to a LAFE. In the theory of such cases, the "plate field" $F_P$ has to be replaced by a different form of macroscopic field, namely the "gap field" $F_G$, and plate FEFs have to be replaced by "gap FEFs". In such cases, the mathematical details are slightly different, but the overall conclusions are essentially the same.

## B. Future solutions

More generally, there seems an urgent need to develop, for each of the possible causes of non-ideality, theory that would allow meaningful characterization parameters to be extracted from measured non-ideal data. In some cases, these meaningful characterization parameters would be those applicable in the limit of very low emission currents, rather than at all current levels.

In the case of non-ideality due to series resistance, in the context of PPP geometry, it is already known how to do this (if the absence of leakage current can be assumed). One writes the following expression for $V_P$

$$V_P = V_m - I_m R_{SR} \,, \qquad (18)$$

where $R_{SR}$ is the (unknown) value of the series resistance. One then makes a FN plot of form $\ln\{I_m/V_P^2\}$ vs $1/V_P$ (or the equivalent MG plot), and evaluates some statistical parameter, called here the "residual", that assesses the linearity of the resulting plot. This is done for a large number of potential values of $R_{SR}$, and the value $R_{SR}^{min}$ that yields the least residual (and hence the "most linear" plot) is identified. Normal characterization-parameter extraction procedures can then be applied to this "most linear" plot. A similar procedure can be applied in other system geometries, and eq. (18) has already been used in several FE contexts[9,14-16].

[In reality, this procedure is not quite correct, because the Murphy-Good FE equation predicts that FN plots are expected theoretically to be slightly curved. A slightly better approach would to carry out a similar procedure with Murphy-Good (MG) plots[13], which are expected to be straight, but it is doubtful whether the resulting numerical improvements would be significant.]

In principle, the needs now are for analogous procedures for all the other potential causes of non-ideality, and for the creation of some "overall statistical comparison parameter" that would enable a decision as to which single cause (or, alternatively, which combination of causes) is most likely to be responsible for the measured data. The creation of comprehensive FE data-analysis theory and procedures of this general kind currently seems many years away. Until then, the best fallback position seems the "third option" described in Section IIIA.

## C. Issues relating to subject research integrity

The author has the subjective impression that the output of reviewed and published papers that report spuriously high FEF values has diminished somewhat in recent years. However, for reasons set out below, it seems important to try to eliminate this source of spurious technological information entirely, as far as is possible. The author's view remains that, unless it is obviously unnecessary, all papers reporting FN-plot-obtained characteristic apex-FEF values, as part of field-emitting-materials characterization, should have had an orthodoxy test applied to their data prior to submission, and should report the outcome. If there is no such report in a FE materials-characterization paper submitted for publication, then reviewers should apply the test themselves, and reject or require major modification of any submission that fails the test.

Elimination of spurious information from FE literature is important in itself, but could also contribute in a wider societal context, given that in recent years some national politicians have appeared to suggest that they consider the scientific reviewing process to be unreliable, at least in some scientific areas. This need for care, in the context of vacuum science and technology, has been stressed in a paper[17] recently published in JVSTA, although problems of the type discussed here were not mentioned.

Reliability problems appear to be more prevalent in biological and medical contexts, and the related "sociology of scientists" has recently been discussed in that context by Charlton[18]. His book takes an extreme view to which the present author does not subscribe, but aspects of his analysis do appear to apply to modern technological FE. My interpretation of his basic argument is that he thinks that science has now become "tribal". "Tribalism" means that researchers take their views from their "tribal in-group", and do not check whether these views are consistent with the thinking of the wider scientific community of which they are part. He considers that, often, the reviewing process has also become "tribalistic". His more basic argument is that part of the societal justification for funding science is that one important scientific objective is to discover and report "truth" for the benefit of society, and that when "scientific tribes" consistently fail to even attempt to do this, then Government (with justification) will eventually remove the funding that supports their activities.

I would encourage my FE colleagues to help as strongly as they can with this business of ensuring the integrity of FE literature, particularly when acting as reviewers.

**Appendix: The FN-plot slope formula**

The formula for interpreting the slope of a FN plot was established by Houston[19] and corrected by Burgess, Kroemer and Houston[20], but is re-derived here because the author is using a new form for the Murphy-Good FE equation.

The Murphy-Good FE equation for $I_m(V_m)$ is conveniently written[13] in the form

$$I_m = A_f a \phi^{-1}(V_m/\zeta_C)^2 \exp[-v_F b \phi^{3/2} \zeta_C/V_m] , \qquad (A1)$$

where $a$ is the first FN constant[7], $A_f$ is the relevant formal area[13], and $v_F$ is the relevant value of the special mathematical function[21] $v(x)$, where $x$ is the "Gauss variable", i.e. the independent variable in the Gauss Hypergeometric Differential Equation. [Due to a change in preferred notation, the symbol $l'$ in ref. 21 should be replaced by $x$.] The value $v_F$ applies to the Schottky-Nordheim (SN) barrier characterized by $\phi$ and $F_C$. It follows that (provided any voltage-dependences in $A_f$ and $\phi$ are weak) the predicted slope $S_V^{\text{theor}}$ of an $I_m(V_m)$-type FN plot is given adequately by:

$$S_V^{\text{theor}} = d\ln\{I_m/V_m^2\}/d(V_m^{-1}) \approx -b\phi^{3/2}\zeta_C\, d\{v_F V_m^{-1}\}/dV_m^{-1}\,. \tag{A2}$$

The final derivative in eq. (A2) is evaluated by: (a) writing $v_F$ explicitly as a function $v(f)$ of the characteristic scaled (barrier) field $f$ defined in this context by

$$f \equiv (e^3/4\pi\varepsilon_0)\,\phi^{-2}\,F_C\,; \tag{A3}$$

and (b) writing $V_m$ in the form:

$$V_m = fV_{mR}\,, \tag{A4}$$

where $V_{mR}$ is the "reference measured-voltage" needed (for an ideal device/system with the same system geometry) to pull the top of a SN barrier down to the Fermi level. The derivative then transforms as follows:

$$d\{v_F V_m^{-1}\}/dV_m^{-1} = d\{v(f)/f\}/df^{-1} = v(f) + f^{-1}dv/df^{-1} = v(f) - fdf = s(f)\,. \tag{A5}$$

The special mathematical function $v(x)$ is applied to FE by setting $x=f$, and it can be recognized that $s(f)$ is a particular value of the special mathematical function $s(x)$ defined by

$$s(x) = v(x) - xdv/dx. \tag{A6}$$

As already noted, MG theory predicts FN plots to be slightly curved. When applying MG theory to the interpretation of FN plots, it is necessary to use the "fitting value" $s_t\,[=s(f_t)]$ (of the slope correction function $s(f)$) that corresponds to the "fitting value" $f_t$ of scaled (barrier) field $f$. This (initially unknown) fitting value $f_t$ is the $f$-value at which the tangent to the theoretical FN plot would be parallel to the straight-line fitted to the experimental data points in a FN plot.

Consequently, the formula for the interpretation of the experimental FN-plot slope $S_V^{\text{fit}}$ becomes

$$\zeta_C^{\text{extr}} = -\,S_V^{\text{fit}}/s_t b\phi^{3/2}\,, \tag{A7}$$

as stated above as eq. (4). The function $s(f)$ is a weakly varying function of $f$, and—as indicated above—it is usually adequate to use the approximation $s_t \approx 0.95$.

[For completeness, we note that a merit of a Murphy-Good (MG) plot[13], as compared with a FN plot, is that the theoretical MG plot is predicted to be "very nearly" straight. Thus, the formula for interpretation of a MG plot does not need to contain a slope correction factor.]

**ACKNOWLEDGMENT**

The author thanks the University of Surrey for provision of office and information facilities.